\documentclass[12pt]{article}
\usepackage{graphicx}
\usepackage{amsmath}
\usepackage{amssymb}

 \newcommand{\R}{{\mathbb R}}

 \newcommand{\somr}{\sum_{r=1}^{k}}

 \newcommand{\ynret}{Y_{r}^{\ast}}

 \newcommand{\inr}{I_{r}}

 \newcommand{\Dnr}{D_{r}}

 \newcommand{\CQFD}
 {
 \mbox{}
 \nolinebreak
 \hfill
 \rule{2mm}{2mm}
 \medbreak
 \par
 }

 \newtheorem{Theo}{Theorem}

 \newtheorem{lemma}{Lemma}
 \newtheorem{Coro}{Corollary}

\begin{document}

\pagestyle{empty}
\font\geant=cmssbx10 at 16pt
\font \nain=cmss10 at 12pt
\begin{center}

\geant LINEAR PROGRAMMING PROBLEMS 
\vskip 10pt
FOR FRONTIER ESTIMATION
\vskip 18pt
\nain G. BOUCHARD, S. GIRARD, A. IOUDITSKI and  A. NAZIN\\
\vskip 20pt
\vskip 20pt

\end{center}

\noindent {\bf Abstract}\\

\noindent We propose new estimates for the frontier of a set of points.
They are defined as kernel estimates
covering all the points and whose associated support is
of smallest surface. The estimates are written as linear combinations
of kernel functions applied to the points of the sample.
The coefficients of the linear combination are then computed by solving
a linear programming problem. In the general case, the solution
of the optimization problem is sparse, that is, only a few coefficients
are non zero. The corresponding points play the role of support vectors
in the statistical learning theory.
The $L_1$ error between the estimated and the true frontiers is
shown to be almost surely converging to zero, and the rate of convergence
is provided. The behaviour of the estimates on finite sample situations
is illustrated on some simulations.\\

\noindent {\bf Contact information}\\

\noindent Guillaume Bouchard: {\sc is}2, INRIA, ZIRST, 655, av. de l'Europe, Montbonnot,\\
38334 Saint-Ismier cedex, France.   {\tt Guillaume.Bouchard@inrialpes.fr}\\

\noindent St\'ephane Girard and Anatoli Iouditski: SMS/LMC, Universit\'e Grenoble I, BP 53,\\
38041 Grenoble cedex 9, France. 
{\tt Stephane.Girard@imag.fr}, {\tt Anatoli.Iouditski@imag.fr} \\

\noindent Alexander Nazin: Institute of Control Sciences, RAS, Profsoyuznaya str., 65, \\
117997 Moscow, Russia.  {\tt nazine@ipu.rssi.ru}\\

\noindent {\bf Acknowledgements}\\

\noindent Financial support from the IAP research network nr P5/24 of the 
Belgian Government (Federal Office for Scientific, Technical and Cultural Affairs) is 
gratefully acknowledged.\\

\noindent The work of A.~Nazin has been carried out during his stay in INRIA Rh\^one-Alpes as
invited professor, November-December 2002.

\section{Introduction}

Many proposals are given in the literature for estimating a set $S$ 
given a finite random set of points drawn from the interior.
This problem of edge or support estimation arises in classification
({\sc Hardy} \& {\sc Rasson}~\cite{HarRas}),
clustering problems ({\sc Hartigan}~\cite{Hart}),
discriminant analysis ({\sc Baufays} \& {\sc Rasson}~\cite{BauRas}),
and outliers detection.  
Applications are found in medical diagnosis ({\sc Tarassenko}
{\it et al}~\cite{THCB}) as well as in condition monitoring of 
machines ({\sc Devroye} \& {\sc Wise}~\cite{DevWis}).
In image analysis,
the segmentation problem can be considered under the support estimation
point of view, where the support is a convex bounded set in $\R^2$
({\sc Korostelev} \& {\sc Tsybakov}~\cite{KorTsy2}).
We also point out some applications in econometrics
(e.g. {\sc Deprins}, {\it et al}~\cite{DepSimTul}).
In such cases, the unknown support can be written
\begin{equation}
\label{defS}
S\triangleq\{(x,y):\;~0\leq~x\leq~1\;~;~\;~0\leq~y~\leq~f(x)\},
\end{equation}
where $f$ is an unknown function. 
Here, the problem reduces to estimating $f$, called the production frontier
(see for instance {\sc H\"ardle} {\it et al}~\cite{Hardle}).
The data consist of pair $(X,Y)$
where $X$ represents the input (labor, energy or capital) used to produce
an output $Y$ in a given firm. In such a framework, the value $f(x)$
 can be interpreted as the maximum level of output which is attainable
for the level of input $x$.
{\sc Korostelev} {\it et al}~\cite{KorTsy}
suppose $f$ to be increasing and concave,
 from economical considerations, which suggests an adapted
estimator, called the DEA (Data Envelopment Analysis) estimator.
It is the lowest concave monotone increasing
function covering all the sample points.
Therefore it is piecewise linear
and, up to our knowledge, it is the first frontier estimate computed thanks to a linear programming technique
({\sc Charnes} {\it et al}~\cite{Charnes}).
Its asymptotic distribution is established by {\sc Gijbels}
{\it et al}~\cite{Gijbels2}.

\noindent 
An early paper was written by {\sc Geffroy}~\cite{Geff1} for 
independent identically distributed observations
from a density $\phi$.
The proposed estimator is a kind of histogram based on the extreme values of the sample.
This work was extended in two main directions.\\

\noindent On the one hand, piecewise polynomials estimates were introduced.  
They are  defined locally on a given slice as the lowest
polynomial of fixed degree covering all the points in the considered slice.
Their optimality in an asymptotic minimax sense is proved under weak
assumptions on the rate of decrease $\alpha$ of the density $\phi$ towards 0
by {\sc Korostelev} \& {\sc Tsybakov}~\cite{KorTsy2} and by
{\sc H\"ardle} {\it et al}~\cite{Hardle2}. Extreme values methods
are then proposed by {\sc Hall} {\it et al}~\cite{Hall} and by
{\sc Gijbels} \& {\sc Peng}~\cite{Gijbels} to estimate the parameter $\alpha$.\\

\noindent
On the other hand, different propositions for smoothing
Geffroy's estimate were made in the case of a Poisson point
process. {\sc Girard} \& {\sc Jacob}~\cite{GirJac3}
 introduced estimates based on kernel regressions
and orthogonal series method~\cite{GirJac,GirJac2}.
In the same spirit, {\sc Gardes}~\cite{Gardes} proposed a Faber-Shauder estimate. 
{\sc Girard} \& {\sc Menneteau}~\cite{GirMen} introduced a general framework
for studying estimates of this type and generalized them to
supports writting 
$$
S=\{(x,y):\;~x\in E\;~;~\;~0\leq~y~\leq~f(x)\},
$$
where $f$ is an unknown function and $E$ an arbitrary set. 
In each case, the limit distribution of the estimator is established. 

\noindent We also refer to {\sc Abbar}~\cite{Abb} and
 {\sc Jacob} \& {\sc Suquet}~\cite{JacSuq} who used a similar
smoothing approach, although their estimates are not based  on
the extreme values of the Poisson process.\\

\noindent The estimate proposed in this paper can
be considered to belong to the intersect of these two directions.  
It is defined as a kernel estimate obtained by smoothing
some selected points of the sample.   These points are
chosen automatically by solving a linear programming problem
to obtain an estimate 
of the support covering all the points and with smallest surface.
Its advantages are the following: it can be computed with
standard optimization algorithms
(see e.g.  {\sc Bonnans} {\it et al}~\cite{Lemarechal}, chapter~4),
its smoothness is directly
linked to the smoothness of the chosen kernel and it benefits
from interesting theoretical properties.  Here, we prove
that it is almost surely convergent for the $L_1$ norm.  
The estimate is defined in Section~\ref{definition}. 
Its theoretical properties are established in Section~\ref{secmain}.
The behaviour of the estimate is illustrated in Section~\ref{simul}
on finite sample situations.   Its compared to a similar
proposition found in {\sc Barron} {\it et al}~\cite{BarBirMas}. 
Proofs are postponed to Section~\ref{proofs}.

\section{Boundary estimates}
\label{definition}

\subsection{A linear programming problem}

Let all the random variables be defined on a probability space
$(\Omega,\mathcal{F},P)$.
The problem under consideration is to estimate an unknown positive
function $f: [0,1]\to (0,\infty)$ on the basis of observations
$Z_N=(X_i,Y_i)_{i=1,\dots,N}$. The former represents an i.i.d.
sequence with pairs
  $(X_i,Y_i)$ being uniformly distributed in the set $S$
  defined as in (\ref{defS}). For the sake of simplicity, we consider
in the following the extension of $f$ on all $\R$ by introducing
$f(x)=0$ for all $x\notin [0,1]$.  
Letting
$$
C_f \triangleq \displaystyle\int_0^1 f(u)\,du=\int_\R f(u)\,du,
$$
each variable $X_i$ is distributed in $[0,1]$ with p.d.f.
$f(\cdot)/C_f$
while $Y_i$ has the uniform conditional distribution with respect to $X_i$
in the interval $[0,f(X_i)]$.
\bigskip

\noindent The considered estimate of the frontier is chosen from the family of 
functions:
\noindent 
\begin{equation}
\label{estimator}
\left\{
\begin{array}{l}
\widehat{f}_N(x) = \displaystyle \sum_{i=1}^N K_h(x-X_i)\alpha_i \,,\qquad
K_h(t)=h^{-1}K(t/h),\\
\alpha_i \geq 0,\qquad i=1,\dots,N,
\end{array}
\right. 
\end{equation}
where $K$ is a kernel function $K:\R\to [0,\infty)$
integrating to one and with bandwidth $h>0$.
 Each coefficient $\alpha_i$ represents
the importance of the point $(X_i,Y_i)$ in the estimation.
In particular, if $\alpha_i\neq 0$, the corresponding point
$(X_i,Y_i)$ can be called a support vector by analogy
with Support Vector Machines (SVM). We refer to {\sc Cristianini} \&
{\sc Shawe-Taylor}~\cite{CriSha} for a review on this topic
and to {\sc Sch\"olkopf} \& {\sc Smola}~\cite{Scho}, chapter~8, for examples
of application of SVM to quantile estimation.
The constraint $\alpha_i\geq 0$ for all $i=1,\dots,N$ ensures that
$\widehat{f}_N(x) \geq 0$ for all $x\in\R$ and prevents the estimator
from being too irregular (see equation (\ref{Lipfunest1})).
Let us remark that the surface of the estimated support is given by
$$
\int_\R \widehat{f}_N(x)\, dx = \sum_{i=1}^N \alpha_i.  
$$
This suggests to define the
vector parameter $\alpha=(\alpha_1,\dots,\alpha_N)^T$
from a linear program as follows
\begin{eqnarray}\label{IPgoal}
J_P^* &\triangleq& \min_\alpha \textbf{1}^T \alpha
\\
\mathrm{subject~to} &&\nonumber
\\
&& A\alpha \geq Y \label{constr1}\\
&&\alpha\geq 0.\label{constr2}
\end{eqnarray}
The following notations have been introduced:
\begin{eqnarray}
\mathbf{1} &\triangleq& (1, 1,\dots, 1)^T\in \R^N \nonumber\\
A &\triangleq& \left\| K_h(X_i-X_j) \right\|_{i,j=1,\dots,N}\nonumber\\
Y &\triangleq& (Y_1,\dots,Y_N)^T.   \nonumber
\end{eqnarray}
Hence, $A\alpha =
(\widehat{f}_N(X_1),\dots,\widehat{f}_N(X_N))^T$, and the vector
constraint (\ref{constr1}) means that
\begin{equation}
\label{vec2scals}
\widehat{f}_N(X_i)\geq Y_i\,,\qquad i=1,\dots,N.  
\end{equation}
In other words, $\widehat{f}_N$ defines the kernel estimate
of the support covering all the points and with smallest surface.
In practice (see Section \ref{simul} for an illustration)
the solution of the linear
program is sparse in the sense that $n(\alpha)=\mbox{Card}\{\alpha_i\neq 0\}$ is small
(for moderate values of $h$) and thus the resulting estimate
is fast to compute even for large samples.  \\

\noindent Let us note that the above described
estimator (\ref{estimator})--(\ref{constr2}) might be derived as
the Maximum Likelihood Estimate related to the approximation family
(\ref{estimator}). Indeed, the joint probability density function
for observations $Z_N$ given parameter function $f(x)$ can be 
written
\begin{equation}
\label{jointpdf}
  p(Z_N\,|\;f) = \prod_{i=1}^N \frac{f(X_i)}{C_f} \cdot \frac{1}{f(X_i)} \mathbf{1}\{0\leq Y_i\leq f(X_i)\},
\end{equation}
where $\mathbf{1}\{.\}$ is the indicator function.  
Moreover,
\begin{equation}\label{Cfhat}
\left.\phantom{\sum} C_{f}^{}\right|_{f=\widehat{f}_N}
=\sum_{i=1}^N \alpha_i,
\end{equation}
and therefore, the Log-Likelihood function is
\begin{equation}\label{LogLihood}
L(\alpha) \triangleq \log p(Z_N\,|\;\widehat{f}_N) =
 -N\log \sum_{i=1}^N \alpha_i
 +\sum_{i=1}^N \log\mathbf{1}\{Y_i\leq \widehat{f}_N(X_i)\},
\end{equation}
and its maximization over the set of non-negative parameters
$\alpha$ is equivalent to problem (\ref{IPgoal})--(\ref{constr2}).

\subsection{Comparison with other methods}

 Let us remark that other solutions for estimating $\alpha$ in 
(\ref{estimator}) have already been proposed.  
{\sc Girard} \& {\sc Menneteau}~\cite{GirMen} considered a partition
 $\{\inr:\;1\leq r\leq k\}$ of $[0,1]$,
with $k\to\infty$. For all $1\leq r\leq k$, they introduce
$ \Dnr=\{ (x,y) : x\in \inr,\;0\leq y\leq f(x)\}$,
the slice of $S$ built on $\inr$, 
$ \ynret=\max\{  Y_i;(X_i,Y_i) \in \Dnr\}$,
and the estimates\\
\begin{center}
$
\widehat{\alpha_i}= 
\left|
\begin{array}{ll}
\lambda(\inr) \ynret & \mbox{ if } \exists\, r\in\{1,\dots,k\}
\,;\, Y_i=\ynret \\ 
0 & \mbox{ otherwise,}
\end{array}
\right.
$
\end{center}
where $\lambda$ is the Lebesgue measure.  
They propose the following frontier estimate
$$
\check{f}_N(x)=\somr  K_h(x-x_r) \lambda(\inr)\ynret,
$$
where $x_r$ is the center of $I_r$.
\noindent This approach suffers from a practical difficulty: the
choice of the partition and more precisely the choice of $k$.
In our context, solving the linear problem (\ref{IPgoal})--(\ref{constr2}) direcly yields the support vectors.\\

\noindent In this sense, the estimate proposed in
{\sc Barron} {\it et al}~\cite{BarBirMas} 
is similar to $\widehat{f}_N$. It is defined by
the Fourier expansion:
\begin{equation}
\widehat{g}_N(x)=c_0 + \sum_{k=1}^M a_k \cos{(2\pi k x)}
+ \sum_{k=1}^M b_k \sin{(2\pi k x )},
\end{equation} 
where the vector of parameters $\beta=(c_0,a_1,\dots,a_M,b_1,\dots,b_M)^T$
is solution of the linear programming problem:
\begin{equation}
\label{objBM}
\min c_0  \quad \left( = \int_0^1 \widehat{g}_N(x)dx\right)
\end{equation} 
under the constraints
\begin{eqnarray}
\label{contBM}
\widehat{g}_N(X_i) & \geq & Y_i,\quad i=1,\dots,N\\
\label{contBM2}
\sum_{k=1}^M k(|a_k| + |b_k|) & \leq & L/(2\pi).
\end{eqnarray}
Therefore, $\widehat{g}_N$ defines the Fourier estimate
of the support covering all the points (equation (\ref{contBM})),
$L$-Lipschitzian (equation (\ref{contBM2})) and
with smallest surface (equation (\ref{objBM})).
From the theoretical point of view, this estimate benefits from
minimax optimality. It is compared to $\widehat{f}_N$
on practical situations in Section \ref{simul} for different
choices of parameters $M$, $L$ and $h$.

\section{Main results}
\label{secmain}

The basic assumptions on the unknown boundary function are:
\begin{itemize}
\item[A1.] $0< f_{\min} \leq f(x) < f_{\max} <\infty$, for all $x\in [0,1]$,
 \item[A2.] $|f(x)-f(y)|\leq L_f |x-y|$, for all $x,y \in [0,1]$, 
$\quad L_f<\infty$.  
\end{itemize}
\medskip


\noindent The following assumptions on the kernel function
are introduced:
\begin{itemize}
  \item[B1.] $K(t)=K(-t)\geq 0$, 
  \item[B2.] $\displaystyle \int_\R K(t)\, dt =1 $,
  \item[B3.] $|K(s)-K(t)|\leq L_K |s-t|$,\, $L_K<\infty$,
  \item[B4.] $C_0(K) \triangleq \displaystyle \int_\R K^2(t)\, dt
 <\infty\,$ and\,
 $C_2(K) \triangleq \displaystyle \int_\R t^2\,K(t)\, dt <\infty$.  
\end{itemize}
We denote $K_{\max}\triangleq\max K(t)$.
In the following theorem the consistency of the estimate
is established with respect to the $L_1$ norm on the
$[0,1]$ interval.

\begin{Theo}\label{Th1}
Let  $h\to 0$ and $\displaystyle{\log{N}/(Nh^2)}\to 0$ as
$N\to\infty$. Let the above mentioned assumptions A and B hold
true. Then estimator (\ref{estimator})--(\ref{constr2}) has the
following asymptotic properties:
\begin{equation}\label{L1rate}
\limsup_{N\to\infty}\, \varepsilon_1^{-1}(N)  \|\hat{f}_N -f\|_{1}
\leq C(\omega)<\infty \quad \mathrm{a.s.}
\end{equation}
with
\begin{equation}\label{epsL1}
  \varepsilon_1(N) \triangleq \max\left\{h,\sqrt{\log{N}/(Nh^2)}\right\}.
\end{equation}
\end{Theo}
\medskip

\begin{Coro}\label{CorTh1}
The maximum rate of convergence which is guaranteed by Theorem
\ref{Th1}
$$
\|\hat{f}_N -f\|_{1} = O \left(\left(\log{N}/N\right)^{1/4}\right)
$$
is attained for
\begin{equation}\label{opth}
  h \asymp \left({\log{N}}/{N}\right)^{1/4}.
\end{equation}
\end{Coro}
\medskip
This rate of convergence can be ameliorated at the price 
of a slight modification of the estimate.  In the following,
an additional constraint is considered in
order to impose to each coefficient $\alpha_i$ to be of order
$1/N$.  The counterpart of this modification is that the
new estimate $\tilde{f}_N$ will usually rely on more support
vectors than $\widehat{f}_N$.  

\noindent Let us modify the estimator
(\ref{estimator})--(\ref{constr2}) as follows.
\begin{equation}\label{Mestimator}
\tilde{f}_N(x) = \sum_{i=1}^N K_h(x-X_i)\alpha_i
\end{equation}
where vector $\alpha=(\alpha_1,\dots,\alpha_N)^T$ is defined from
the Modified Linear Program
\begin{eqnarray}\label{MIPgoal}
J_{MP}^*&\triangleq&\min_\alpha \textbf{1}^T \alpha
\\
\mathrm{subject~to} &&\nonumber
\\
&& A\alpha \geq Y \label{Mconstr1}\\
&&0\leq\alpha\leq {C_\alpha}/{N} \label{Mconstr2}
\end{eqnarray}
with a constant
\begin{equation}\label{C_alpha}
   C_\alpha >f_{\max}.
\end{equation}
\medskip

\noindent\textbf{Remark.} In fact, we need to ensure $C_\alpha >
C_f$ which is implied by (\ref{C_alpha}). \bigskip


\noindent The modified estimator
(\ref{Mestimator})--(\ref{C_alpha}) differs
from that of (\ref{estimator})--(\ref{constr2}) by additionally
bounding each
 $\alpha_i$ from above, see constraints (\ref{Mconstr2}).
 Below we prove that under condition (\ref{C_alpha})
 as well as finite support kernel $K(\cdot)$
the Modified Linear Program (\ref{MIPgoal})--(\ref{Mconstr2}) has
a nonempty set of admissible solutions with the same upper bound
as (\ref{UB}) and a better lower bound than (\ref{LB}).

\begin{Theo}\label{MTh1}
Let  $h\to 0$ and $\displaystyle{\log{N}/(Nh)}\to 0$ as
$N\to\infty$. Let
 kernel function $K(\cdot)$ has a finite support, that is
 $K(t)=0\;\forall\, |t|\geq 1$,
 and the assumptions A and B hold true.
 Then estimator (\ref{Mestimator})--(\ref{C_alpha}) has the
following asymptotic properties:
\begin{equation}\label{ML1rate}
\limsup_{N\to\infty}\, \varepsilon_2^{-1}(N)  \|\tilde{f}_N -f\|_{1}
\leq C(\omega)<\infty \quad \mathrm{a.s.}
\end{equation}
with
\begin{equation}\label{MepsL1}
  \varepsilon_2(N) \triangleq \max\left\{h,\sqrt{\log{N}/(Nh)}\right\}.
\end{equation}
\end{Theo}
\medskip

\noindent\textbf{Remark.} The support of $K(\cdot)$ is fixed to be
the interval $[-1,1]$ without loss of generality.

\medskip

\begin{Coro}\label{MCorTh1}
The maximum rate of convergence which is guaranteed by Theorem
\ref{MTh1}
$$
\|\tilde{f}_N -f\|_{1} = O \left(\left(\log{N}/N\right)^{1/3}\right)
$$
is attained for
\begin{equation}\label{Mopth}
  h \asymp \left({\log{N}}/{N}\right)^{1/3}.
\end{equation}
\end{Coro}
\medskip

\section{Numerical experiments}
\label{simul}

The simulations presented here illustrate the behaviour of the kernel estimator $\widehat{f}_N$ compared to the estimator based on Fourier expansions
$\widehat{g}_N$ proposed in {\sc Barron} {\it et al}~\cite{BarBirMas}. 
Since the Fourier estimator $\widehat{g}_N$ 
requires the unknown function to be periodic,
we choose $f$ such that $f(0)=f(1)$.
Besides, to avoid boundary effects on the input domain,
we consider functions that are nearly zero when $x$ 
is close to 0 or 1. In more general situation, boundary
corrections should be implemented 
(see {\sc Cowling} \& {\sc Hall}~\cite{Cow}).
The chosen function
\begin{eqnarray*}
f(x) = 0.1 	&+& 5(x-0.1){\mathbf 1}_{\{x>0.1\}}\\ 
			&-& 5(x-0.2){\mathbf 1}_{\{x>0.2\}}\\  
			&+& 1(x-0.5){\mathbf 1}_{\{x>0.5\}} \\ 
			&-& 9(x-0.8){\mathbf 1}_{\{x>0.8\}} \\ 
			&+& 8(x-0.9){\mathbf 1}_{\{x>0.9\}}, 
\end{eqnarray*}
is piecewise linear and locally Lipschitizian with a Lipschitz constant
$L_f=8$.
For each estimate, the $L_1$ error $\Delta_N$ 
as well as the number of effective parameters $np$
(that is $n_\alpha$ and  $n_\beta=\mbox{Card}\{\beta_i\neq 0\}$) are evaluated
for $N=25$ and $N=100$.
The average value and the standard deviation of these quantities are
computed on 1000 replications in the first case and on 100 replications
in the second one.
The estimation is carried out with different values of the parameters,
namely $h$ for the kernel estimate, and $L$ and $M$ for the Fourier estimate.
The adaptive choice of these parameters is not implemented in this setting. 
The results are summarized in Tables~\ref{simul25} and \ref{simul102}.
The lowest error is emphasized for each estimate.
It can be noted that the mean $L_1$ error of both estimates are very similar.
In fact, the kernel estimate seems to give a slight lower error for
small number of points and the Fourier estimate yields better results 
for large sample size situations, confirming its asymptotic optimality.
Let us note that the standard deviation of the $L_1$ error is in
general smaller for the kernel estimate.
Regarding the number of parameters, the kernel estimate seems to 
be more parsimonious than the Fourier estimate.


\begin{table}[p]

\begin{tabular}{|c|ccc|cc|cc|}
\hline
estimate &	$h$ &	$L$ &	$M$ &	mean($\Delta_N$) &	st-dev($\Delta_N$) &	mean($np$) &	st-dev($np$)\\
\hline
kernel &	 0.100 &	  &	  &	 0.123 &	 0.038 &	 5.263 &	 0.970\\
&	 0.120 &	  &	  &	 0.116 &	 0.034 &	 4.490 &	 0.841\\
&	 \textbf{0.140} &	  &	  &	 \textbf{0.112} &	 \textbf{0.033} &	 \textbf{3.841} &	 \textbf{0.683}\\
&	 0.160 &	  &	  &	 0.115 &	 0.031 &	 3.420 &	 0.636\\
&	 0.180 &	  &	  &	 0.123 &	 0.027 &	 3.120 &	 0.657\\
&	 0.200 &	  &	  &	 0.132 &	 0.023 &	 2.863 &	 0.645\\
\hline
Fourier &	  &	 3.000 &	 4.000 &	 0.144 &	 0.035 &	 4.567 &	 0.777\\
&	  &	 \textbf{5.000} &	 \textbf{4.000} &	 \textbf{0.119} &	 \textbf{0.043} &	 \textbf{5.508} &	 \textbf{0.986}\\
&	  &	 7.000 &	 4.000 &	 0.127 &	 0.043 &	 6.572 &	 1.217\\
&	  &	 9.000 &	 4.000 &	 0.138 &	 0.044 &	 7.235 &	 1.284\\
&	  &	11.000 &	 4.000 &	 0.147 &	 0.046 &	 7.592 &	 1.249\\
&	  &	13.000 &	 4.000 &	 0.154 &	 0.046 &	 7.815 &	 1.210\\
\hline
Fourier &	  &	 3.000 &	 8.000 &	 0.144 &	 0.036 &	 4.581 &	 0.800\\
&	  &	 \textbf{5.000} &	 \textbf{8.000} &	 \textbf{0.121} &	 \textbf{0.044} &	 \textbf{5.571} &	 \textbf{1.057}\\
&	  &	 7.000 &	 8.000 &	 0.129 &	 0.044 &	 6.730 &	 1.379\\
&	  &	 9.000 &	 8.000 &	 0.142 &	 0.046 &	 7.632 &	 1.669\\
&	  &	11.000 &	 8.000 &	 0.153 &	 0.047 &	 8.314 &	 1.873\\
&	  &	13.000 &	 8.000 &	 0.163 &	 0.048 &	 8.859 &	 2.050\\
\hline
\end{tabular}
\caption{\label{simul25}Results for 1000 simulations with $N=25$ points.}
\end{table}


\begin{table}[p]

\begin{tabular}{|c|ccc|cc|cc|}
\hline
estimate &	$h$ &	$L$ &	$M$ &	mean($\Delta_N$) &	st-dev($\Delta_N$) &	mean($np$) &	st-dev($np$)\\
\hline
kernel &	 0.050 &	  &	  &	 0.073 &	 0.016 &	13.700 &	 1.560\\
&	 0.070 &	  &	  &	 0.060 &	 0.014 &	 9.890 &	 1.246\\
&	 \textbf{0.090} &	  &	  &	 \textbf{0.060} &	 \textbf{0.013} &	 \textbf{7.350} &	 \textbf{1.132}\\
&	 0.110 &	  &	  &	 0.063 &	 0.012 &	 5.820 &	 0.989\\
&	 0.130 &	  &	  &	 0.075 &	 0.012 &	 4.690 &	 0.734\\
&	 0.150 &	  &	  &	 0.085 &	 0.013 &	 3.960 &	 0.549\\
\hline
Fourier &	  &	 3.000 &	 4.000 &	 0.129 &	 0.021 &	 5.120 &	 0.700\\
&	  &	 5.000 &	 4.000 &	 0.078 &	 0.020 &	 5.790 &	 0.756\\
&	  &	 \textbf{7.000} &	 \textbf{4.000} &	 \textbf{0.061} &	 \textbf{0.012} &	 \textbf{7.630} &	 \textbf{0.960}\\
&	  &	 9.000 &	 4.000 &	 0.064 &	 0.013 &	 8.700 &	 0.560\\
&	  &	11.000 &	 4.000 &	 0.069 &	 0.015 &	 8.880 &	 0.409\\
&	  &	13.000 &	 4.000 &	 0.071 &	 0.016 &	 8.950 &	 0.297\\
\hline
Fourier &	  &	 3.000 &	 8.000 &	 0.129 &	 0.021 &	 5.160 &	 0.762\\
&	  &	 5.000 &	 8.000 &	 0.078 &	 0.020 &	 5.920 &	 0.849\\
&	  &	 7.000 &	 8.000 &	 0.059 &	 0.013 &	 8.070 &	 1.350\\
&	  &	 \textbf{9.000} &	 \textbf{8.000} &	 \textbf{0.059} &	 \textbf{0.015} &	\textbf{10.470} &	 \textbf{1.630}\\
&	  &	11.000 &	 8.000 &	 0.063 &	 0.015 &	12.090 &	 1.682\\
 &	  &	13.000 &	 8.000 &	 0.069 &	 0.015 &	13.620 &	 2.068\\
\hline
\end{tabular}
\caption{\label{simul102}Results for 100 simulations with $N=100$ points.}
\end{table}

\section{Proofs}
\label{proofs}

The proof of Theorem \ref{Th1} which is given in subsection
\ref{PrTh1} is based on both upper and lower bounds derived below.

\subsection{Upper bound for $\widehat{f}_N$}

\begin{lemma}\label{Lm1Th1}
Let  $h\to 0$ and $\displaystyle{\log{N}/(Nh)}\to 0$ as
$N\to\infty$. Let the above mentioned assumptions A and B hold
true. Then for almost all $\omega\in\Omega$ there exist finite
number $N_0(\omega)$ such that
\begin{equation}\label{UB}
  J^*_P \leq C_f + O(h)
+ O\left(\sqrt{{\log{N}}/(Nh)}\;\right),
     \qquad\forall\; N\geq N_0(\omega),
\end{equation}
with non random both $O(h)$ and
$O\left(\sqrt{\log{N}/(Nh)}\;\right)$.
\end{lemma}
\medskip

\noindent\textbf{Proof of Lemma \ref{Lm1Th1}.} 
\medskip\noindent 1.  Since kernel function $K(.)$ is supposed to be even then matrix
$A$ is symmetric, and the dual problem associated 
to (\ref{IPgoal}) -- (\ref{constr2}) can be written:
\begin{eqnarray}\label{IDgoal}
J_D^*&\triangleq&\max_\lambda Y^T\lambda
\\
\mathrm{subject~to} &&\nonumber
\\
&& A\lambda \leq \mathbf{1} \label{IDconstr}\\
&&\lambda\geq 0.
\end{eqnarray}
Let us replace vector $Y$ in (\ref{IDgoal}) for
\begin{equation}
F \triangleq(f(X_1),\dots,f(X_N))^T
\end{equation}
and, moreover, change the vector constraint (\ref{IDconstr}) by a
scalar one which is directly obtained by just summing all $N$ rows
of (\ref{IDconstr}). Thus, we arrive at the modified dual problem
\begin{eqnarray}\label{IDMgoal}
J_{MD}^*&\triangleq&\max_\lambda F^T\lambda
\\
\mathrm{subject~to} &&\nonumber
\\
&&  \mathbf{1}^T A\lambda \leq N \label{IMDconstr} \\
&&\lambda\geq 0. \label{lamplus}
\end{eqnarray}
Since $F \geq Y$ and according to the well known Duality Theorem
(see e.g.  {\sc Hiriart-Urruty} \& {\sc Lemar\'echal}~\cite{HUL},
chapter 7):
\begin{equation}\label{res1}
J_P^* = J_D^* \leq J_{MD}^*.
\end{equation}
Now we derive an upper bound on $J_{MD}^*$\,.
\bigskip

\noindent 2. Let us arbitrarily fix a vector $\lambda$ which meet the
constraints (\ref{IMDconstr}), (\ref{lamplus}) and then write
inequality (\ref{IMDconstr}) in the equivalent form as follows:
\begin{equation}
\frac{1}{N} \sum_{j=1}^N \lambda_j \left( K_h(0) + \sum_{i\neq
j}^N K_h(X_i-X_j) \right) \leq 1,
\end{equation}
or, equivalently,
\begin{equation}\label{inegal}
\frac{1}{N} \sum_{j=1}^N \lambda_j \left( \frac{1}{h} K(0) +
\sum_{i\neq j}^N E\left\{ K_h(X_i-X_j) \;|\,  X_j \right\} +
\sum_{i\neq j}^N \xi_{ij} \right) \leq 1,
\end{equation}
with
$$
  \xi_{ij} \triangleq K_h(X_i-X_j)
- E\left\{ K_h(X_i-X_j) \;|\,  X_j\right\}.
$$
Now apply upper bound (\ref{Log1}), proved in Lemma \ref{AuxLm1}
(see Appendix),
 to the relation (\ref{inegal}) taking into account that
$K(0)>0$ and
  \begin{eqnarray} \label{EKXj}
E\left\{ K_h(X_i-X_j) \;|\,  X_j \right\} &=& \frac{1}{h} \int_0^1
K\left( \frac{u-X_j}{h} \right)  \frac{f(u)}{C_f}\, du
\\ \nonumber
&=& \frac{1}{C_f} \int_\R K(t)  f(X_j+ht)\, dt
\\
&=& \frac{1}{C_f} \left( f(X_j) +O(h) \right),
 \label{EKXj2}
  \end{eqnarray}
  with non random $O(h)$.
Hence,
\begin{equation}\label{inegal2}
\frac{N-1}{C_{f}N} \sum_{j=1}^N \lambda_j \left( f(X_j) +O(h) -
        C\sqrt{\frac{\log{N}}{Nh}}\;\right)
 \leq 1,
     \qquad\forall\; N\geq N_2(\omega),
\end{equation}
 with non random constant $C$.
First, inequality (\ref{inegal2}) implies
\begin{equation}\label{sumlam}
\sum_{j=1}^N \lambda_j \leq \frac{2C_f}{f_{\min}}<\infty,
     \qquad\forall\; N\geq N_3(\omega),
\end{equation}
with almost surely finite $N_3(\omega)\geq N_2(\omega)$.
Second, (\ref{sumlam}) and (\ref{inegal2}) imply
 upper bound (\ref{UB}) and  Lemma \ref{Lm1Th1} is proved.
\CQFD

\subsection{Lower bound for $\widehat{f}_N$}
\begin{lemma}\label{Lm2Th1}
Under the assumptions of Theorem \ref{Th1},  for almost all
$\omega\in\Omega$ there exist finite number $N_1(\omega)$ such
that for each $x\in (0,1)$
\begin{equation}\label{LB}
  \widehat{f}_N(x) \geq f(x)- O\left(\sqrt{{\log{N}}/{(N h^2)}}\;\right),
     \qquad\forall\; N\geq N_1(\omega),
\end{equation}
where $O(\cdot)$ do not depend on $x$.
\end{lemma}
\medskip

\noindent\textbf{Proof of Lemma \ref{Lm2Th1}.}
\medskip\noindent 1. Suppose that for some non-random $\delta_x>0$ there exists 
(with probability one) an integer
$i_k\in\{1,\dots,N\}$ such that
\begin{equation}\label{epsuppose}
 |x-X_{i_k}| \leq \delta_x.
\end{equation}
Then, the estimation error at a point $x\in(0,1)$
can be expanded as 
\begin{eqnarray}\label{ferr1}
  f(x)-\widehat{f}_N(x) &=& \left[f(x)-f(X_{i_k})\right]\\
 \label{ferr2}      &+& \left[f(X_{i_k})- \widehat{f}_N(X_{i_k})\right]\\
 \label{ferr3}      &+& \left[\widehat{f}_N(X_{i_k}) - \widehat{f}_N(x)\right].
\end{eqnarray}
The term in the right hand side (\ref{ferr1}) may be bounded as
follows
\begin{equation}\label{err1}
  \left|f(x)-f(X_{i_k})\right| \leq L_f  \left|x-X_{i_k} \right| \leq L_f \delta_x,
\end{equation}
as well as the term (\ref{ferr3})
\begin{equation}\label{err3}
  \left|\widehat{f}_N(X_{i_k}) - \widehat{f}_N(x)\right|
  \leq L_{\widehat{f}_N}  \left|x-X_{i_k} \right| \leq L_{\widehat{f}_N}\, \delta_x,
\end{equation}
with a Lipschitz constant $L_{\widehat{f}_N}\;$ for the function
estimate $\widehat{f}_N(x)$, which is bounded below.
In order to bound (\ref{ferr2}) assume that for some non-random $\delta_y>0$,
\begin{equation}\label{deltay}
  Y_{i_k}\geq f(X_{i_k}) -\delta_y \mbox{ a.s}.
\end{equation}
Remind that $\widehat{f}_N(X_{i_k})\geq Y_{i_k}$ due to
(\ref{constr1}) or (\ref{vec2scals}). Thus,
\begin{equation}\label{err2}
f(X_{i_k})- \widehat{f}_N(X_{i_k}) \leq (Y_{i_k}+\delta_y) -
Y_{i_k} =\delta_y.
\end{equation}
Combining all these bounds we obtain from (\ref{ferr1}) that for
all $N\geq N_0(\omega)$,
\begin{equation}\label{ferr12}
f(x)-\widehat{f}_N(x) \leq \delta_y + \left(L_f +
L_{\widehat{f}_N} \right) \delta_x.
\end{equation}
 \medskip \noindent  2. Note that a straightforward evaluation of the  Lipschitz constant for the estimate function yields:
\begin{eqnarray}\label{Lipfunest1}
 |\widehat{f}_N(u) - \widehat{f}_N(v)| &\leq&
\sum_{i=1}^N \alpha_i \left|K_h(u-X_i)-K_h(v-X_i) \right|
\\ \label{Lipfunest2}
&\leq& \frac{L_K}{h^2} \left( \sum_{i=1}^N \alpha_i \right) |u-v|.
\end{eqnarray}
Hence, due to the upper bound (\ref{UB}), we obtain almost surely
\begin{equation}\label{Lipfunest}
  L_{\widehat{f}_N} = \frac{L_K}{h^2}\, C_f (1+o(1)),
      \qquad\forall\; N\geq N_0(\omega),
\end{equation}
with almost surely finite $N_0(\omega)$.\\
 \medskip \noindent  3. Now, we demonstrate that under appropriate definition of $\delta_x$ and $\delta_y$ as functions of $h$ and $N$ there exist almost
surely finite random integer $N_0(\omega)$ such that
\begin{equation}\label{xikyik}
  \forall\, N\geq N_0(\omega),\qquad
  \exists\, i_k\in \{1,\dots,N\, : \; \left( X_{i_k},Y_{i_k}\right)\in \Delta(x) \},
\end{equation}
with
\begin{equation}\label{Deltaxset}
\Delta(x) \triangleq \{(u,v)\,:\; |x-u|\leq \delta_x, \,
f(x)-\delta_y\leq v\leq f(u)\}.
\end{equation}
Indeed, introduce
\begin{equation}\label{dely}
\delta_y \triangleq\left(\frac{\kappa\log{N}}{N h^2}\right)^{1/2},
\end{equation}
and
\begin{equation}\label{delxh2dely}
  \delta_x = h^2\delta_y.
\end{equation}
Then,
\begin{eqnarray}\nonumber
    P\{(X_i,Y_i)\notin\Delta(x)\quad\forall\,i=1,\dots,N\}
    &=& \left( 1 -\frac{1+o(1)}{C_f}\delta_x\delta_y\right)^N
\\
    &=& \left( 1-\frac{1+o(1)}{C_f}h^2\delta_y^2\right)^N
\nonumber\\
 & \leq & \exp\left\{ -\frac{1+o(1)}{C_f}N h^2 \delta_y^2\right\}
\nonumber\\
\label{probinDelx} &\leq& N^{-\kappa/(2C_f)}.
  \end{eqnarray}
Hence, fixing
\begin{equation}\label{kappa2}
\kappa > 2C_f
\end{equation}
implies the convergence of the series
\begin{equation}\label{series}
  \sum_{N=1}^\infty P\{(X_i,Y_i)\notin\Delta(x)\quad\forall\,i=1,\dots,N\} <\infty,
\end{equation}
which, due to Borel--Cantelly lemma, implies the existence of almost
surely finite $N_0(\omega)$ such that relation (\ref{xikyik})
holds true.\\
 \medskip \noindent  4. Therefore, substituing
 relations (\ref{Lipfunest}), (\ref{dely}),
and (\ref{delxh2dely}) to (\ref{ferr12}) leads to lower bound
\begin{eqnarray}\nonumber
\widehat{f}_N(x) &\geq& f(x)- \delta_y - O\left( h^{-2} \right)
\delta_x \\
&=& f(x)- O\left(\sqrt{\frac{\log{N}}{N h^2}}\;\right),
\label{LBestim}
\end{eqnarray}
with non-random term $O(\cdot)$ independent of $x$.
\CQFD

\subsection{Proof of Theorem \ref{Th1}}\label{PrTh1}
 \medskip \noindent  1. Since $|u|=u-2u\mathbf{1}\{u<0\}$,
 the $L_1$-norm of estimation error can be expanded as
\begin{eqnarray}\label{L1norm1}
  \|\hat{f}_N -f\|_{1} &=& \int_0^1 \left[\hat{f}_N(x) -f(x)\right]\,dx
\\ \label{L1norm2}
&& +2\int_0^1 \left[f(x)-\hat{f}_N(x) \right] \mathbf{1}\! \left\{\hat{f}_N(x)<f(x)\right\}\,dx.  
\end{eqnarray}
 \medskip \noindent  2.  Applying Lemma \ref{Lm1Th1} to the right hand side (\ref{L1norm1}) yields
\begin{equation}\label{useUB}
  \limsup_{N\to\infty}\, \varepsilon_{UB}^{-1}(N)
    \left(\int_0^1 \left[\hat{f}_N(x) -f(x)\right]\,dx\right)
    \leq \mathrm{const}<\infty
\quad\mathrm{a.s.}
\end{equation}
with
\begin{equation}\label{eps1N}
  \varepsilon_{UB}(N) \triangleq \max\left\{h,\sqrt{\log{N}/(Nh)}\right\}.
\end{equation}
 \medskip \noindent  3. In order to obtain a similar result for the term (\ref{L1norm2}), note that Lemma \ref{Lm2Th1}
implies
$$
\zeta_N(x) \triangleq \varepsilon_{LB}^{-1}(N) \left[f(x)-\hat{f}_N(x) \right]
\leq C(\omega)<\infty \quad \mathrm{a.s.}
$$
uniformly with respect to both $x$ and $N$,
with
\begin{equation}\label{epsLB}
  \varepsilon_{LB}(N) \triangleq \sqrt{\log{N}/(Nh^2)}.
\end{equation}
Hence, one may apply Fatou lemma, taking into account that
$u\mathbf{1}\{u>0\}$ is a continuous, monotone function:
\begin{eqnarray}
  &&  \limsup_{N\to\infty}\, \varepsilon_{LB}^{-1}(N)
  \int_0^1 \left[f(x)-\hat{f}_N(x) \right] \mathbf{1}\! \left\{\hat{f}_N(x)<f(x)\right\}\,dx
   \\
 &\leq&
 \int_0^1 \limsup_{N\to\infty}\, \zeta_N(x)\, \mathbf{1}\!
    \left\{\zeta_N(x)>0\right\}\,dx
   \\
 &\leq& C(\omega)<\infty \quad \mathrm{a.s.}
\end{eqnarray}
\medskip 
\noindent 4. Thus, the obtained relations together with (\ref{L1norm1}) and
(\ref{L1norm2}) imply (\ref{L1rate}), (\ref{epsL1}) and
Theorem \ref{Th1} is proved. \CQFD

\noindent The proof of Theorem \ref{MTh1} which is given in subsection
\ref{MPrTh1} is based on the similar ideas as that of Theorem
\ref{Th1}, see below.

\subsection{Upper bound for $\tilde{f}_N$}

Since the admissible set (\ref{Mconstr1}), (\ref{Mconstr2}) is
narrower being compared to that of (\ref{constr1}),
(\ref{constr2}), it is important to demonstrate that the upper
bound remains at least the same.
\begin{lemma}\label{MLm1Th1}
Let the assumptions of Theorem \ref{MTh1} hold true. Then for
almost all $\omega\in\Omega$ there exist finite number
$N_0(\omega)$ such that
\begin{equation}\label{MUB}
  J^*_{MP} \leq C_f + O(h)
+ O\left(\sqrt{\frac{\log{N}}{Nh}}\;\right),
     \qquad\forall\; N\geq N_0(\omega),
\end{equation}
with non random both $O(h)$ and
$O\left(\sqrt{\log{N}/(Nh)}\;\right)$.
\end{lemma}
\medskip

\noindent\textbf{Proof of Lemma \ref{MLm1Th1}.}
 \medskip\noindent 1.
Since kernel function $K(t)$ is supposed to be even then matrix
$A$ is symmetric, and the related to
(\ref{MIPgoal})--(\ref{Mconstr2}) dual problem looks like
\begin{eqnarray}\label{MIDgoal}
J_{MD}^*&\triangleq&\max_{\lambda,\,\nu}\left(Y^T\lambda -C_\alpha
N^{-1} \mathbf{1}^T\nu \right)
\\
\mathrm{subject~to} &&\nonumber
\\
&& A\lambda -\nu \leq \mathbf{1} \label{MIDconstr}\\
&&\lambda\geq 0\\
 &&\nu\geq 0.
\end{eqnarray}
Let us replace vector $Y$ in (\ref{MIDgoal}) for
\begin{equation}
F \triangleq(f(X_1),\dots,f(X_N))^T,
\end{equation}
and, moreover, change the vector constraint (\ref{MIDconstr}) by a
scalar one which is directly obtained by just summing all $N$ rows
of (\ref{MIDconstr}). Thus we arrive at the modified dual problem
\begin{eqnarray}\label{MIDMgoal}
J_{MMD}^*&\triangleq&\max_{\lambda,\,\nu} \left(F^T\lambda
-C_\alpha N^{-1} \mathbf{1}^T\nu \right)
\\
\mathrm{subject~to} &&\nonumber
\\
&&  \mathbf{1}^T A\lambda -\mathbf{1}^T \nu \leq N \label{MIMDconstr} \\
&&\lambda\geq 0 \label{Mlamplus}\\
 &&\nu\geq 0. \label{Mnuplus}
\end{eqnarray}
Since $F \geq Y$ and according to the well known Duality Theorem
\begin{equation}\label{Mres1}
J_{MP}^* = J_{MD}^* \leq J_{MMD}^*.
\end{equation}
Now, we derive an upper bound on $J_{MMD}^*$\,.
\bigskip

\noindent  2. Let us arbitrarily fix $(\lambda, \nu)$  which meet the
constraints (\ref{MIMDconstr})--(\ref{Mnuplus}) and then write
inequality (\ref{MIMDconstr}) in the equivalent form as follows:
\begin{equation}
\frac{1}{N} \sum_{j=1}^N \lambda_j \left( K_h(0) + \sum_{i\neq
j}^N K_h(X_i-X_j) \right) \leq 1 +\frac{1}{N} \mathbf{1}^T\nu,
\end{equation}
or, equivalently,
\begin{equation}\label{Minegal}
\frac{1}{N} \sum_{j=1}^N \lambda_j \left( \frac{1}{h} K(0) +
\sum_{i\neq j}^N E\left\{ K_h(X_i-X_j) \;|\,  X_j \right\} +
\sum_{i\neq j}^N \xi_{ij} \right) \leq 1 +\frac{1}{N} \mathbf{1}^T\nu,
\end{equation}
with
\begin{equation}
  \xi_{ij} \triangleq K_h(X_i-X_j)
- E\left\{ K_h(X_i-X_j) \;|\,  X_j\right\}.
\end{equation}
Now apply upper bound (\ref{Log1}), proved in Lemma \ref{AuxLm1},
 to the relation (\ref{Minegal}) taking into account that
$K(0)>0$ as well as (\ref{EKXj})--(\ref{EKXj2}). 
Hence,
\begin{equation}\label{Minegal2}
\frac{N-1}{C_{f}N} \sum_{j=1}^N \lambda_j \left( f(X_j) +O(h) -
        C\sqrt{\frac{\log{N}}{Nh}}\;\right)
 \leq 1 +\frac{1}{N}\mathbf{1}^T\nu,
     \qquad\forall\; N\geq N_2(\omega),
\end{equation}
 with non random constant $C$.
First, from inequality (\ref{Minegal2}) it follows that
\begin{equation}\label{Msumlam}
\sum_{j=1}^N \lambda_j \leq \frac{C_f}{f_{\min}}\left(2 +\frac{1}{N}
                            \mathbf{1}^T\nu\right),
     \qquad\forall\; N\geq N_3(\omega),
\end{equation}
with almost surely finite $N_3(\omega)\geq N_2(\omega)$.
Consequently, as it follows from (\ref{Minegal2}), for almost all
$\omega\in\Omega$ and sufficiently large $N$
\begin{eqnarray}\label{Minegal4}
F^T\lambda - \frac{C_\alpha}{N} \mathbf{1}^T\nu &\leq& C_f
\left(1+ O(h) + O\left(\sqrt{\frac{\log{N}}{Nh}}\;\right)\right)
\\ \label{Minegal4n}
&& -(C_\alpha -C_f(1+o(1)))\mathbf{1}^T\nu,
\end{eqnarray}
 with non random $O\left(\sqrt{\log{N}/(Nh)}\;\right)$.
Thus, (\ref{Mres1}) and (\ref{Minegal4}) prove the upper bound
(\ref{MUB}), since (\ref{Mconstr2}) implies $C_\alpha>C_f$.
\CQFD

\subsection{Lower bound for $\tilde{f}_N$}

\begin{lemma}\label{MLm2Th1}
Under the assumptions of Theorem \ref{MTh1},  for almost all
$\omega\in\Omega$ there exist finite number $N_1(\omega)$ such
that for each $x\in (0,1)$
\begin{equation}\label{MLB}
  \tilde{f}_N(x) \geq f(x)- O\left(\sqrt{{\log{N}}/(N h})\;\right),
     \qquad\forall\; N\geq N_1(\omega),
\end{equation}
where $O(\cdot)$ do not depend on $x$.
\end{lemma}
\medskip

\noindent\textbf{Proof of Lemma \ref{MLm2Th1}} is given in the
same manner as that of Lemma \ref{Lm2Th1}. The only essential
difference is in better Lipschitz constant for $\tilde{f}_N(x)$.
Indeed, for any $u, v\in (0,1)$
\begin{eqnarray}\label{MLipfunest1}
 \left|\tilde{f}_N(u) - \tilde{f}_N(v)\right| &\leq&
\sum_{i=1}^N \alpha_i  \left|K_h(u-X_i)-K_h(v-X_i) \right|
\\ \label{MLipfunest2}
&\leq& \frac{L_K}{h^2} \left(\sum_{i\in I(u)} \alpha_i +\sum_{i\in
I(v)} \alpha_i \right) |u-v|,
\end{eqnarray}
with
\begin{equation}\label{Iuv}
    I(\cdot) \triangleq \{i\,|\; K_h(\cdot-X_i)\neq 0 \}.
\end{equation}
From the Strong Law of Large Numbers,
\begin{equation}\label{cardIuv}
    \mathrm{Card}\, I(\cdot) =\frac{f(\cdot)}{C_f}\,
    Nh(1+o(1))\quad\mathrm{a.s.}
\end{equation}
and thus,
%
\begin{equation}\label{MLipfunest}
  L_{\tilde{f}_N} = \frac{L_K}{h^2}\, \frac{C_\alpha}{N}\, \frac{2f_{\max}}{C_f}\, N h
  = O\left(\frac{1}{h}\right)
\end{equation}
by the upper bound (\ref{Mconstr2}) on $\alpha$.
\CQFD

\subsection{Proof of Theorem \ref{MTh1}}\label{MPrTh1}
Theorem \ref{MTh1} is proved in the same manner as that of Theorem
\ref{Th1}, basing on lemmas \ref{Lm1Th1} and \ref{Lm2Th1}. Note, that
the lower bound from Lemma \ref{Lm2Th1} is now not worse being
compared to the upper bound, which is the result of the estimator
modification.\\

\noindent\textbf{Note:} The result (\ref{ML1rate})--(\ref{MepsL1})
of Theorem \ref{MTh1} may also be proved for differentiable kernel
functions with infinite support which meet the condition
\begin{equation}\label{kernelcondi}
    \left|K'(t)\right| \leq \mu K(t), \quad \forall\, t\in \R,
\end{equation}
with some constant $\mu$. Indeed, (\ref{kernelcondi}) implies
\begin{equation}\label{kerconimpl}
\left|\tilde{f}_N^{\;\prime}(x)\right| \leq \frac{1}{h^2}
\sum_{i=1}^N \alpha_i \left|K'\left(\frac{x-X_i}{h}\right)\right|
\leq \frac{\mu}{h} \tilde{f}_N(x).
\end{equation}
Consequently, when estimate function $\tilde{f}_N(x)$ is bounded
from above, its Lipschitz constant is of order
$O\left(h^{-1}\right)$ that is the same as in (\ref{MLipfunest}).

\section{Appendix}

\begin{lemma}\label{AuxLm1}
Let the assumptions A and B hold true and
 constant $C$ be sufficiently large. Define the random variables
\begin{equation}
  \xi_{ij} \triangleq K_h(X_i-X_j)
- E\left\{ K_h(X_i-X_j) \;|\,  X_j\right\}, \quad i\neq j.
\end{equation}
Then,
 for almost all $\omega\in\Omega$ there
exist finite integer $N_2(\omega)$ such that
\begin{equation}\label{Log1}
\max_{j=1,\dots,N} \left|\,\frac{1}{N-1}\sum_{i\neq j}^N
    \xi_{ij}\, \right|\leq\, C \sqrt{{\log{N}}/(Nh)}
    \qquad\forall\; N\geq N_2(\omega).
\end{equation}
\end{lemma}
\medskip

\noindent\textbf{Proof of Lemma \ref{AuxLm1}.} Note that for each
$j=1,\dots,N$ the unbiased i.i.d. random variables
$\left(\xi_{ij}\right)|_{i\neq j}$ have the following properties:
\begin{equation}
 \left| \xi_{ij} \right| \leq
\frac{2}{h}K_{\max} \triangleq a,
\end{equation}
and
\begin{eqnarray}\label{sigma1}
  E\left\{ \xi_{ij}^2  \;|\,  X_j \right\}
&\leq& \frac{1}{h^2 C_f}\int_0^1 K^2\left(\frac{u-X_j}{h} \right)
f(u)\,du
 \nonumber\\
&\leq& \frac{1}{h C_f}\int_\R K^2(t)\, f(X_j+ht)\, dt
 \nonumber\\
&\leq& \frac{C_0(K)}{h C_f}f_{\max} \triangleq\sigma_1^2.
\end{eqnarray}
Thus, one may apply the Bernstein inequality (see, e.g., 
{\sc Birg\'e} \& {\sc Massart}~\cite{BirMas} or {\sc Bosq}~\cite{Bosq},
Theorem 2.6) which leads to
  \begin{eqnarray}
   P\left\{\left|\,\frac{1}{N-1}\sum_{i\neq j}^N
    \xi_{ij}\, \right|>\,\mu  \quad\left|\;  X_j \phantom{\frac{1}{h}}\!\!\! \right.
    \right\}
    &\leq & 2\exp\left(-\frac{(N-1)\mu^2}{2(\sigma_1^2 + a\mu/3)} \right).
    \nonumber
  \end{eqnarray}
Let us put
\begin{equation}
  \mu = \sqrt{\frac{\kappa\log{N}}{Nh}},
\end{equation}
with sufficiently large $\kappa$ which is defined below. Hence,
for all $N\geq N_1$,  $N_1$ being sufficiently large non random
integer,
  \begin{eqnarray}
   P\left\{\left|\,\frac{1}{N-1}\sum_{i\neq j}^N
    \xi_{ij}\, \right|>\,\sqrt{\frac{\kappa\log{N}}{Nh}}
     \quad\left|\;  X_j \phantom{\frac{1}{h}}\!\!\! \right.
    \right\}
    &\leq & 2 N^{-\kappa_1},
    \nonumber
  \end{eqnarray}
with
\begin{equation}
  \kappa_1 \triangleq \frac{\kappa\, C_{f}f_{\max}}{3C_0(K)}.
\end{equation}
Therefore,
\begin{eqnarray}
  && P\left\{\max_{j=1,N} \left|\,\frac{1}{N-1}\sum_{i\neq j}^N
    \xi_{ij}\, \right|>\,\sqrt{\frac{\kappa\log{N}}{Nh}}
     \quad\left|\;  X_j \phantom{\frac{1}{h}}\!\!\! \right.
    \right\}
     \\
  &\leq & \sum_{j=1}^N P\left\{ \left|\,\frac{1}{N-1}\sum_{i\neq j}^N
    \xi_{ij}\, \right|>\,\sqrt{\frac{\kappa\log{N}}{Nh}}
     \quad\left|\;  X_j \phantom{\frac{1}{h}}\!\!\! \right.
    \right\}
    \\
  &\leq& 2 N^{1-\kappa_1}.
\end{eqnarray}
Consequently, any fixed parameter
\begin{equation}\label{kappa}
\kappa >\frac{6C_0(K)}{C_{f}f_{\max}}
\end{equation}
ensures $\kappa_1>2$ which implies the convergence of series
$\displaystyle\sum^\infty N^{1-\kappa_1}$ and, due to
Borel--Cantelli lemma, the desired result (\ref{Log1}).
\CQFD


\newpage


\begin{thebibliography}{xx}

\bibitem{Abb} { Abbar, H.} (1990)
 { Un estimateur spline du contour d'une r\'epartition ponctuelle al\'eatoire.}
 {\em Statistique et analyse des donn\'ees}, {\bf 15}(3), 1--19.

\bibitem{BarBirMas} Barron, A.R., Birg\'e, L. and Massart, P. (1999) Risk Bounds for model selection via penalization. 
\textit{Probab.  Theory Relat. Fields}, \textbf{113}, 301--413.

\bibitem{BauRas}{ Baufays, P. and Rasson, J.P.} (1985)
{ A new geometric discriminant rule.}
{\em Computational Statistics Quaterly}, {\bf 2}, 15--30.

\bibitem{BirMas} Birg\'e, L. and Massart, P. (1995) 
Minimum contrast estimators on sieves. 
{\em Preprint Universit\'e Paris Sud, France}, {\bf 95-42}.

\bibitem{Lemarechal} Bonnans, F., Gilbert, J.C.,  Lemar\'echal, C.
and Sagastiz\'abal, C. (1997) 
Optimisation num\'erique. Aspects th\'eoriques et pratiques.
in {\em Math\'ematiques \& Applications}, {\bf 27}, Springer, Paris.

\bibitem{Bosq} { Bosq, D.} (2000)
{Linear processes in function spaces. Theory and applications.} 
in {\em Lecture Notes in Statistics}, {\bf 149}, Springer-Verlag, New York.

\bibitem{Charnes} Charnes, A., Cooper, W.W. and Rhodes, E. (1978)
Measuring the inefficiency of decision making units.
{\it European Journal of Operational Research}, {\bf 2}, 429--444.

\bibitem{Cow} Cowling, A. and  Hall, P. (1996)
On pseudodata methods for removing boundary effects in kernel density estimation. 
{\it Journal of the Royal Statistical Society B}, {\bf 58}, 551--563.

\bibitem{CriSha} Cristianini, N. and Shawe-Taylor, J.  (2000)
{\it An introduction to support vector machines},
Cambridge University Press.  

\bibitem{DepSimTul}{ Deprins, D., Simar, L. and Tulkens, H.} (1984)
{ Measuring Labor Efficiency in Post Offices.} in
{\em The Performance of Public Enterprises: Concepts and Measurements} by
M. Marchand, P. Pestieau and H. Tulkens, North Holland ed, Amsterdam.

\bibitem{DevWis}{ Devroye, L.P. and Wise, G.L.} (1980)
{ Detection of abnormal behavior via non parametric estimation of the support.}
{\em SIAM J. Applied Math.}, {\bf 38}, 448--480.

\bibitem{Gardes} {Gardes, L.} (2002)
{ Estimating the support of a Poisson process via the {F}aber-{S}hauder
basis and extreme values.}
{\em Publications de l'Institut de Statistique de l'Universit\'e de Paris},
 {\bf XXXXVI}, 43--72.

\bibitem{Geff1} { Geffroy, J.} (1964)
{ Sur un probl\`eme d'estimation g\'eom\'etrique.}
{\em Publications de l'Institut de Statistique de l'Universit\'e de Paris},
 {\bf XIII}, 191--200.

\bibitem{Gijbels} {Gijbels, I. and Peng, L.} (1999).
{ Estimation of a support curve via order statistics.}
{\em Discussion Paper} {\bf 9905}, Institut de Statistique,
Universit\'e Catholique de Louvain.

\bibitem{Gijbels2} { Gijbels, I., Mammen, E., Park, B.U. and Simar, L.} (1999).
{On estimation of monotone and concave frontier functions.}
{\em Journal of the American Statistical Association}, {\bf 94}, 220--228. 

\bibitem{GirJac} { Girard, S. and Jacob, P.} (2002a)
{ Extreme values and Haar series estimates of point processes boundaries.}
{\em Scandinavian Journal of Statistics}, to appear.

\bibitem{GirJac2} { Girard, S. and Jacob, P.} (2002b)
{ Projection estimates of point processes boundaries.}
{\em Journal of Statistical Planning and Inference}, to appear.

\bibitem{GirJac3} { Girard, S. and  Jacob, P.} (2001)
{ Extreme values and kernel estimates of point processes boundaries.}
{\em Technical report ENSAM-INRA-UM2}, {\bf 01-02}.

\bibitem{GirMen} { Girard, S. and  Menneteau, L.} (2002)
{ Limit theorems for extreme values estimates of point processes boundaries.}
{\em Technical report INRIA}, {\bf RR-4366}.

\bibitem{Hall} {Hall, P., Nussbaum, M. and Stern, S.E.} (1997)
{On the estimation of a support curve of indeterminate sharpness.} 
{\em J. Multivariate Anal.}, {\bf 62}, 204--232.

\bibitem{Hardle}{ H\"ardle, W., Hall, P. and Simar, L.} (1995)
{ Iterated boostrap with application to frontier models.}
{\em J. Productivity Anal.}, {\bf 6}, 63--76.

\bibitem{Hardle2}{ H\"ardle, W., Park, B. U. and Tsybakov, A. B.} (1995)
{ Estimation of a non sharp support boundaries.}
{\em J. Multiv. Analysis}, {\bf 43}, 205--218.

\bibitem{Hardle3}{H\"ardle, W.} (1990)
{\em Applied nonparametric regression},
Cambridge University Press, Cambridge.

\bibitem{HarRas}{ Hardy, A. and Rasson, J.P.} (1982)
{ Une nouvelle approche des probl\`emes de classification automatique.}
{\em Statistique et Analyse des donn\'ees}, {\bf 7}, 41--56.

\bibitem{Hart}{ Hartigan, J.A.} (1975)
{\em Clustering Algorithms,} Wiley, Chichester.

\bibitem{HUL} Hiriart-Urruty, J.B., Lemar\'echal, C. (1993)
Convex analysis and minimization algorithms. Part 1: Fundamentals. 
in {\em Grundlehren der Mathematischen Wissenschaften}, {\bf 305},
Springer-Verlag, Berlin.

\bibitem{JacAbb} { Jacob, P. and Abbar, H.} (1989)
{ Estimating the edge of Cox process area.}
{\em Cahiers du Centre d'Etudes de Recherche Op\'erationnelle}, {\bf 31},
 215--226.

\bibitem{JacSuq} { Jacob, P. and Suquet, P.} (1995)
{ Estimating the edge of a Poisson process by orthogonal series.}
{\em Journal of Statistical Planning and Inference}, {\bf 46},
 215--234.

\bibitem{KorTsy}{ Korostelev, A., Simar, L. and Tsybakov, A. B.} (1995)
{Efficient estimation of monotone boundaries.}
{\em The Annals of Statistics}, {\bf 23}, 476--489.

\bibitem{KorTsy2}{  Korostelev, A.P. and Tsybakov, A.B.} (1993)
{ Minimax theory of image reconstruction.}
in {\em Lecture Notes in Statistics}, {\bf 82}, Springer-Verlag, New York.

\bibitem{Scho}{Sch\"olkopf, B. and Smola, A. } (2002)
{\em Learning with kernels,} MIT University Press, Cambridge.

\bibitem{THCB}{ Tarssenko, L., Hayton, P., Cerneaz, N. and Brady, M.} (1995)
Novelty detection for the identification of masses in mammograms.
In {\em Proceedings fourth IEE International Conference on Artificial
Neural Networks}, 442--447, Cambridge.

\end{thebibliography}
\end{document}